\documentclass[%
superscriptaddress,
preprint,
showpacs,preprintnumbers,
 amsmath,amssymb,
 aps,
 prl,
]{revtex4-1}

\usepackage{graphicx}% Include figure files
\usepackage{color}% color
\usepackage{dcolumn}% Align table columns on decimal point
\usepackage{bm}% bold math
\usepackage{hyperref}% add hypertext capabilities
\usepackage{CJK}

\begin{document}
%\linenumbers
\begin{CJK*}{GBK}{}
%\preprint{APS/123-QED}

\title{Simultaneous Magnetic and Charge Doping of Topological Insulators with Carbon}%

\author{Lei Shen}%
\email{shenlei@nus.edu.sg}
\affiliation {Department of Physics, 2 Science drive 3, National University of Singapore, Singapore 117542, Singapore}%

\author{Minggang Zeng}%
\affiliation {Department of Physics, 2 Science drive 3, National University of Singapore, Singapore 117542, Singapore}%

\author{Yunhao Lu}%
\affiliation {Department of Materials Science and Engineering, Zhejiang University, Hangzhou 310027, China}%

\author{Ming Yang}%
\affiliation {Department of Physics, 2 Science drive 3, National University of Singapore, Singapore 117542, Singapore}%

\author{Yuan Ping Feng}%
\email{phyfyp@nus.edu.sg}%
\affiliation {Department of Physics, 2 Science drive 3, National University of Singapore, Singapore 117542, Singapore}%

\begin{abstract}
A two-step doping process, magnetic followed by charge or vice versa, is required to produce insulating massive surface states in topological insulators for many physics and device applications. Using first-principles calculations, we demonstrate here simultaneous magnetic and hole doping achieved with a single dopant, carbon, in Bi$_2$Se$_3$. Carbon substitution for Se (C$_{\textrm{Se}}$) results in an opening of a sizable surface Dirac gap (53-85~meV), while the Fermi level ($E_\textrm{F}$) remains inside the bulk gap and close to the Dirac point at moderate doping concentrations. The strong localization of 2$p$ states of C$_{\textrm{Se}}$ favors spontaneous spin polarization via a $p$-$p$ interaction and formation of ordered magnetic moments mediated by the surface state. Meanwhile, holes are introduced into the system by C$_{\textrm{Se}}$. This dual function of carbon doping suggests a simple way to realize insulating massive topological surface states.

\end{abstract}


\maketitle
\end{CJK*}

The second generation 3D topological insulators (TIs), Bi$_2$Se$_3$ and Bi$_2$Te$_3$, are a class of time-reversal-invariant materials characterized by an insulating bulk state and conducting surface state consisting of a single Dirac cone at the $\Gamma$ point (see Refs. \cite{Qi2011RMP,Hasan2010RMP} and references therein). Such massless surface states are protected by the time-reversal symmetry (TRS) and are immune to surface disorders such as defects \cite{Liu2009PRL,Park2010PRL,Hsieh2009PRL}. Many striking physics and device applications of TIs have been proposed, such as quantum anomalous Hall effect \cite{Yu2010Science,Chang2013Science}, magnetic monopole imaging \cite{Qi2009Science}, topological contribution to the Faraday and Kerr effects \cite{Qi2008PRB}, inverse spin-galvanic effect \cite{Garate2010PRL}, optical injection of spins \cite{Lu2010PRB} and TI-based $p$-$n$ junctions \cite{Wray2011arxiv,Wang2012PRB-2}. To enable the above applications, it is necessary to open a surface energy gap as well as keep the Fermi energy inside the bulk gap\cite{Qi2008PRB}. It was proposed that doping TIs by magnetic transition metals (TM), such as Fe, Mn, Cr and Co, could break the TRS and open a surface gap\cite{Liu2009PRL}, which were commonly used experimentally now \cite{Chen2010Science,Wray2010NP,Okada2011PRL,West2012PRB, Hor2010PRB,Henk2012PRL-2,Liu2012PRL,Ye2011arxiv,Yu2010Science,Chang2013Science,Chang2013AM}. %{The resulting magnetic moments are out-of-plane and have a long-range order which can be mediated by surface states or carriers \cite{Hor2010PRB,Liu2009PRL,Wray2010NP,Scholz2012PRL,Chang2013AM,Chang2013Science,Ye2011arxiv}. If the Fermi level ($E_\textrm{F}$) is near the Dirac point, any ordered magnetism is able to open a surface energy gap \cite{Scholz2012PRL,Chen2010Science,Wray2010NP}.
In the experiment, the existence of anionic vacancies in the as-growth Bi$_2$Se$_3$ or Bi$_2$Te$_3$ thin film lead to a typical $n$-type material \cite{Xu2012arxiv-4,Wray2010NP,Chen2010Science,Zhang2011NC}. To move the Fermi level inside the bulk gap from the bulk conduction band, the excess electrons must be compensated (via hole doping) which can be achieved by doping with divalent cations (Mg$^{2+}$ or Ca$^{2+}$) \cite{Chen2010Science}, or chemical molecules with strong electron accepting ability (NO$_2$, CO) \cite{Xu2012arxiv-4,Wray2010NP}. Alloying, such as ternary Bi$_{0.08}$Sb$_{1.92}$Te$_3$ \cite{Zhang2011NC,Chang2013AM,Chang2013Science} and quaternary  Bi$_{1.5}$Sb$_{0.5}$Te$_{1.8}$Se$_{1.2}$ \cite{Xia2013PRB}, is another possible approach to compensate the excess electrons but it is difficult to accurately control the compositions of these compounds\cite{Chang2013Science,Chang2013AM,Xia2013PRB}. Nevertheless, it remains a challenge to find a reliable yet simple scenario to produce out-of-plane magnetization, a stable long-range magnetic order and proximity of Fermi level to the Dirac point. These 3 conditions are necessary in order to experimentally observe the Dirac gap and quantum anomalous Hall effect using the approach of magnetic doping \cite{Liu2009PRL,Abanin2011PRL,Biswas2010PRB,Valla2012PRL,Honolka2012PRL,Scholz2012PRL,Chang2013AM,Chang2013Science}.

The spontaneous spin polarization and local moments in carbon-doped ZnO, a diluted magnetic semiconductor (DMS), has been successfully demonstrated both experimentally and theoretically \cite{Pan2007PRL,Peng2009PRL}. Since O and Se have similar electronic configurations and similar chemical properties, and C has a quite similar electronegativity (2.55) but two more valance holes compared to Se, one can expect that substitution of C for Se would compensate both the Se vacancies and excess electrons in the as-growth Bi$_2$Se$_3$. This motivated us to consider carbon doping as a mean to achieve massive topological surface states (TSS) in Bi$_2$Se$_3$.

In this Letter, we report results of our investigation on the topological surface state of carbon doped Bi$_2$Se$_3$ using first-principles electronic structure calculations. It is found that substitution of C for Se in Bi$_2$Se$_3$ introduces local moments as well as holes. Carbon doping can lead to simultaneous opening of the Dirac gap up to 85~meV and pinning of the Fermi level inside the bulk energy gap. It is striking that magnetic doping and ordering can be achieved using a nonmagnetic dopant (C). We discuss the origin of magnetic moment, mechanism of magnetic coupling, and cause of magnetocrystalline anisotropy in carbon doped TIs system, and compare them with TM-doped TIs and C-doped DMSs. The effects of carrier and impurity concentration on the magnetic state and Dirac gap are also discussed.

First-principles calculations have been widely used to predict topological
insulators with great success\cite{Zhang2013Pss}. Our First-principles calculations were carried out using the \textsc{vasp} code \cite{VASP1} with the projector-augmented-wave potentials and the Perdew-Burke-Ernzerhof generalized gradient approximation \cite{PBE1} for electron exchange-correlation functional. The experimental lattice constants of Bi$_2$Se$_3$ were used in our calculations \cite{crystal}. The Bi$_2$Se$_3$ thin film was modeled by a slab of six quintuple Se-terminated layers (QLs). A vacuum layer of more than 30~\AA~thick was included to ensure a negligible interaction between neighboring slabs \cite{Xu2012arxiv-4,Chen2011PRL}. All structures were relaxed until the force on each atom is smaller than 0.01~eV/\AA~. The cutoff energy was taken to be 400 eV and $k$-point meshes of $7\times 7\times 1$, $5\times 5\times 1$ and $3\times 3\times 1$ were used for the $2\times 2$, $3\times 3$, and $5\times 5$ surface unit cell, respectively. Different surface unit cells were used to study effect of doping concentration on the magnetic and electronic properties. The SOI was included in the topological surface state calculations.

Since Se vacancies are easily formed in Bi$_2$Se$_3$ (Bi-rich growth condition experimentally) \cite{Xia2009NP,Xu2012arxiv-4,Chen2010Science,Zhang2010NP}, we first considered a Se vacancy (V$_{\textrm{Se}}$) on the Bi$_2$Se$_3$ surface which was modeled by removing a Se atom from the top quintuple layer in the supercell. Due to the layered structure of Bi$_2$Se$_3$, there are three inequivalent V$_{\textrm{Se}}$ sites, with V$_{\textrm{Se}}$ at the top (1), middle (2) and bottom (3) of the Bi$_2$Se$_3$ layer, as shown in Fig.~1a. The calculated vacancy formation energies are 0.661~eV, 0.475~eV and 0.158~eV for sites 1-3, respectively. The vacancy formation energy is defined as $E = E_{\rm def} - E_{\rm prin} + n\mu_\textrm{Se}$, where $E_{\rm def}$ and $E_{\rm prin}$ are total energies of the film with and without the Se vacancy, $\mu_{\textrm{Se}}$ is the chemical potential of atomic Se, and $n$ is number of Se vacancies in the supercell. In order to simulate the experimental Bi-rich growth condition \cite{Xia2009NP,Xu2012arxiv-4,Chen2010Science,Zhang2010NP}, the chemical potential of atomic Bi ($\mu_{\textrm{Bi}}$) was taken to be the total energy per atom in Bi bulk phase, while $\mu_{\textrm{Se}}$ is obtained from $2\mu_\textrm{Bi} + 3\mu_\textrm{Se} = E_{{\rm Bi}_2{\rm Se}_3}$, with $E_{{\rm Bi}_2{\rm Se}_3}$ being the total energy of bulk Bi$_2$Se$_3$ per molecular formula. The calculated formation energies indicate that Se vacancy is more likely to form at site~3 in Fig.~1a. The band structure of the 6QL  Bi$_2$Se$_3$ thin film with a single Se vacancy at site~3 is shown in Fig.~1b. As can be seen, the topological surface state with the Dirac-cone characteristics is preserved within the bulk band gap in the presence of the Se vacancy, but the Fermi level $E_F$ is located in the bulk conduction band (BCB), almost 0.5 eV above the Dirac point, due to the anionic Se vacancy. This is in good agreement with experimental observations (inset of Fig.~1b) \cite{Zhang2010NP}.

We next consider filling of the Se vacancy by carbon. Theoretically, it should not be difficult to fill the Se vacancy in Bi$_2$Se$_3$ with carbon. This is because carbon has the same electronegativity (2.55) as Se, but much smaller atomic radius (70 pm). Furthermore, the large interstitial space in the Bi$_2$Se$_3$ crystal lattice, due to the large covalent radius of Bi (146) and Se (116), is able to facilitate diffusion of carbon to V$_\textrm{Se}$ sites on the Bi$_2$Se$_3$ surface. As the doping concentration increases, carbon can further replace selenium during high temperature annealing \cite{West2012PRB}. Figure~1c shows the band structure of Bi$_2$Se$_3$ in the same configuration as in Fig.~1b but the V$_\textrm{Se}$ is filled by a carbon, resulting in a carbon substitutional doping. It shows that carbon doping leads to the opening of a sizable surface Dirac gap (53~meV), which is comparable to that in Fe or Mn doped Bi$_2$Se$_3$ \cite{Chen2010Science,Wray2010NP}. Such a large Dirac gap is desirable for room temperature device applications. In addition, the Fermi level is drawn closer to the surface gap from the bulk conduction band due to hole doping introduced by C$^{4-}_{\textrm{Se}^{2-}}$ (see Fig.~1a). This indicates that we can realize insulating massive TSS in one step, in contrast to the two-step process by TM and carrier doping. Experimentally, a one-step doping procedure would be much preferred over the two-step process because it means easy control of experimental environment and parameters.

In order to understand how a nonmagnetic element such as carbon can open a Dirac gap and tune the Fermi level simultaneously, we calculated the projected density of states (DOS) of carbon doped Bi$_2$Se$_3$ thin film and present the results in Fig.~2. A strong coupling between the 2$p$ states of the dopant (C) and the 4$p$/6$p$ states of the host (Se/Bi) is clearly seen at the valence band maximum (VBM) and conduction band maximum (CBM). This $p$-$p$ interaction is essentially a result of quantum-mechanical level repulsion, which splits the majority and minority 2$p$ states of carbon and ``pushes" the minority 2$p$ states up inside the bulk band gap. The exchange splitting energy between the majority and minority spin $t_2$ states at the zone center [$\Delta\epsilon_\Gamma = \epsilon(t^\uparrow_2) - \epsilon(t^\downarrow_2)$] is substantial ($\sim$0.80~eV). Further examination on the projected $p$ states of carbon [see inset of Fig.~2] shows the $p_x$ and $p_y$ orbitals are degenerate. The spin-down $p_{x,y}$ states are partially occupied, while the spin-down $p_{z}$ states are completely empty. The energy difference ($\Delta\epsilon_\Pi$) between $p^\downarrow_{x,y}$ and $p^\downarrow_{z}$ is around 0.66~eV. Such a splitting is due to the change of carbon bonding environment, especially in the $z$ direction, under the tetrahedral crystal field. The symmetry and wave function of the impurity state ($p$-like $t_2$) are similar to those of the VBM and CBM of Bi$_2$Se$_3$ which consists mainly of anion Se 4$p$ and cation Bi 6$p$ orbitals (without SOI) [see Fig.~2] or cation 6$p$ and anion 4$p$ orbitals (with SOI) due to band inversion. Therefore, a strong $p$-$p$ coupling between the impurity state and valence-band state is allowed near the Fermi level, resulting in the spin polarization and formation of local moments. Such local moments are quite stable and the spontaneous spin-polarization energy is 175~meV. Our calculations also show (see discussion on \textit{impurity concentration effect} below) that these local moments favor an out-of-plane long-range ferromagnetic coupling on the surface of Bi$_2$Se$_3$, mediated by the surface state. The carbon valence orbitals favor more stable fourfold ($\Gamma=a_1+t_2$) splitting under the tetrahedral crystal field, forming C$^{4-}$ (see Fig.~1), which provides two more holes compared to Se$^{2-}$. Therefore, C$_{\rm Se}$ can introduce holes as well as local moments. This duality of carbon doping leads to the simultaneous topological surface state and Fermi level tuning in this intrinsic nonmagnetic topological system. In topological insulators, light elements are often used to tune the surface hole doping by modifying their concentration\cite{Weng2011PRB}. Experimentally, the carrier and impurity density are two key parameters. We therefore discuss their effects on the stability and Fermi level tuning in carbon-doped Bi$_2$Se$_3$.

\textit{Carrier effect ---} Carriers play critical roles in magnetically doped TIs, in both bulk and surface. They mediate the long-range magnetic order \cite{Hor2010PRB} and control the Fermi level as well \cite{Wray2010NP,Chen2010Science}. Electrons in the system also affect the localization of magnetic moments, especially in systems with delocalized $p$ band and hole-induced magnetic system. Here, we consider only the effect of electrons on the local moments and the Dirac gap in C-doped Bi$_2$Se$_3$. As shown in Fig.~3, C substitutional doping at the anion site induces a magnetic moment of $\sim$2$\mu_B$ per anion. If excess electrons are introduced into the system which is possible in an experimental environment, they would quench the magnetic moments. the Dirac gap can be closed if the carrier concentration is sufficiently high. We wish to point out, however, in our calculation the additional electrons are assumed to be homogenously distributed in the whole system, rather than localized at or around the carbon dopants.

\textit{Impurity concentration effect ---}The impurity concentration can affect both the degree of localization of magnetic moments and the hole density because the carbon dopants introduce local moments as well as holes into the system. Figure~4 shows the dependence of the local moment and Dirac gap on the impurity concentration which is given in terms of distance between impurities, by assuming a uniform dopant distribution. It is noted that the inter-impurity distance in the figure is discontinuous. A high impurity concentration enhances the magnetic coupling and introduces more holes, but too large an impurity concentration would quench the local moment and reduce the size of the Dirac gap. This can be understood based on the phenomenological band-coupling model \cite{Dalpian2006SSC}. Under a very low impurity concentration, the impurity states are localized and isolated, which is insufficient to open a Dirac gap. At a moderate doping concentration, the 2$p$ states with the same spin couple to each other but remains localized (FM-I phase), and a Dirac gap opens. The localized impurity bands become broader and delocalized (FM-II phase) as the impurity concentration increases further. This eventually leads to a charge transfer from the majority spin state to the minority spin state, and a reduced local moment and exchange splitting (Fig.~4), and a corresponding reduction of the Dirac gap. If the impurity concentration is sufficiently high, the much reduced exchange splitting would lead to an AFM ground state (AFM phase), even though the energy gain of the FM phase is usually larger than that of the AFM phase in C-doped TI due to the second-order nature of the superexchange interaction of the AFM phase. Therefore, a moderate carbon doping concentration is necessary to open a sizable Dirac gap and to tune the Fermi level close to the Dirac point.

\textit{C-doped TIs vs. TM-doped TIs ---} There are several important differences between the C-doped and Fe/Mn-doped TIs due to the distinctly different orbitals (2$p$ vs. 3$d$) of the dopants. First, the $p$ orbitals of C are usually fully occupied in ionic states, leaving no room for unpaired electrons compared to the $d$ orbitals of transition metals. Second, the spin-orbit and hyperfine interactions in C are considerably weaker compared to that in TMs since they scale as the fourth power of the atomic number. Consequently, it is possible to preserve spin-coherence over time and distance in C-doped TIs much longer than in TM-doped TIs, especially for the \emph{massive} particles in the presence of a Dirac gap. Finally, concerning the mechanism of magnetic ordering in TIs, Liu $et~al.,$ proposed mediation of the long-range ferromagnetism by surface states through the RKKY interaction in TM-doped TIs \cite{Abanin2011PRL,Biswas2010PRB,Liu2009PRL}, while Chang $et~al.$ suggested the $p$-$d$ van Vleck mechanism (localized $p$ valence electrons) for the long-range magnetic order in TM-doped TIs \cite{Chang2013AM}. Clearly, the ferromagnetism in carbon doped Bi$_2$Se$_3$ challenges our understanding of magnetic ordering in such systems because there are no $d$ states near either conductance or valence bands in C-doped TI system (see Fig.~2). The above $s$-$d$ RKKY or $p$-$d$ van Vleck mechanism cannot be applied here. The ferromagnetic double-exchange mechanism can produce large spin moments, but it is a short-range interaction that requires mixed valence, {\em i.e.},  $2p^n\leftrightarrow 2p^{n+1}$. However, there is no evidence that mixed valence occurs in C-doped Bi$_2$Se$_3$ based on our magnetic moment calculations. We propose that the spontaneous spin polarization is induced by the $p$-$p$ interaction and the magnetic exchange coupling in such a system is mediated by the surface states \cite{Liu2009PRL,Chang2013AM}. Further investigation is required to clarify the magnetic exchange in C-doped Bi$_2$Se$_3$ as well as TIs doped by other $2p$ light elements in general.

\textit{C-doped TIs vs. C-doped DMSs ---} We also wish to point out an important difference between magnetism in C-doped TIs and C-doped DMSs, which is essentially due to the different atomic orbitals in the host materials ($3d$ in DMS vs. $4f$ in TIs). Bismuth surface shows a strong SOI due to its high atomic numbers \cite{Koroteev2004PRL}. A large  orbital moment is observed on the surface \cite{Koroteev2004PRL}, which is driven by a localized spin through SOI. Such an orbital moment gives rise to an effective field that is responsible for the large surface magnetic anisotropy in C-doped Bi$_2$Se$_3$ (around 10~meV per C), which is comparable to that in Fe-doped Bi$_2$Se$_3$ \cite{Wray2010NP} and Co-doped Bi$_2$Se$_3$ \cite{Schmidt2011PRB}. The magnetic order in C-doped TIs can be expected to survive up to room temperature since thermal energy is insufficient to destroy such a magnetic anisotropy.  In contrast, the SOI in DMSs is much weaker since the host materials are typically GaAs, ZnO $et~al.$ Therefore, the Curie temperature of most DMSs is below room temperature despite a few exceptions such as C-doped ZnO \cite{Pan2007PRL}. C-doped TIs and C-doped DMSs are also different in their dependence on carriers. The insulating FM phase of C-doped TIs can be carrier-independent for the QAH effect and spin Hall device applications \cite{Chang2013AM}. This is different in C-doped DMSs where carriers (hole) of certain density are indispensable for the long-range FM coupling between local moments and for spin field-effect transistor (FET) applications \cite{Pan2007PRL}.

In summary, we propose carbon substitutional doping at Se site as a {\em one-step} approach to achieve massive topological surface states in Bi$_2$Se$_3$, which is more effective compared to the usual approach of magnetic impurity doping followed by chemical charge doping or vice versa. This is possible because carbon doping simultaneously introduces localized spin moments and holes. Such spin moments are long-range ordering by $p$-$p$ interaction. Meanwhile, holes introduced by carbon doping pull the Fermi level inside the bulk gap from the bulk conduction band. We expect this simple approach of producing massive topological surface states to promote experimental studies in fabricating C-doped TIs or similar materials, and eventually lead to simpler experimental procedure and robust TIs for broader device applications, such as CMOS-like, spin logic and next-generation low-power devices utilizing the charge on/off states (because of opened gap), spin up/down states (because of topological surface) and Hall conductance (because of anomalous Hall effect).

%\textit{Note added in proof.---}During our manuscript preparing, we note a similar work, in which opening Dirac gap of Bi$_2$Te$_3$ by N doping is proposed \cite{Niu2012APL}.

Authors thank Shou-Cheng Zhang and Qi-Kun Xue for their insightful discussions and valuable inputs. This work was carried out using the CCMS Supercomputing System at Institute for Materials Research, Tohoku University, funded by the National Research Foundation Singapore under its Competitive Research Programme (CRP Award No. NRF-CRP 8-2011-06).

%\bibliography{TI}

%\begin{thebibliography}{xx}%
%\bibitem{CNT}H. J. Choi, J. Ihm, S. G. Louie, and M. L. Cohen, Phys. Rev. Lett. {\bf 84}, 2917 (2000).
%\end{thebibliography}%


\clearpage

\begin{figure}
\centering
\includegraphics[width=0.90\textwidth]{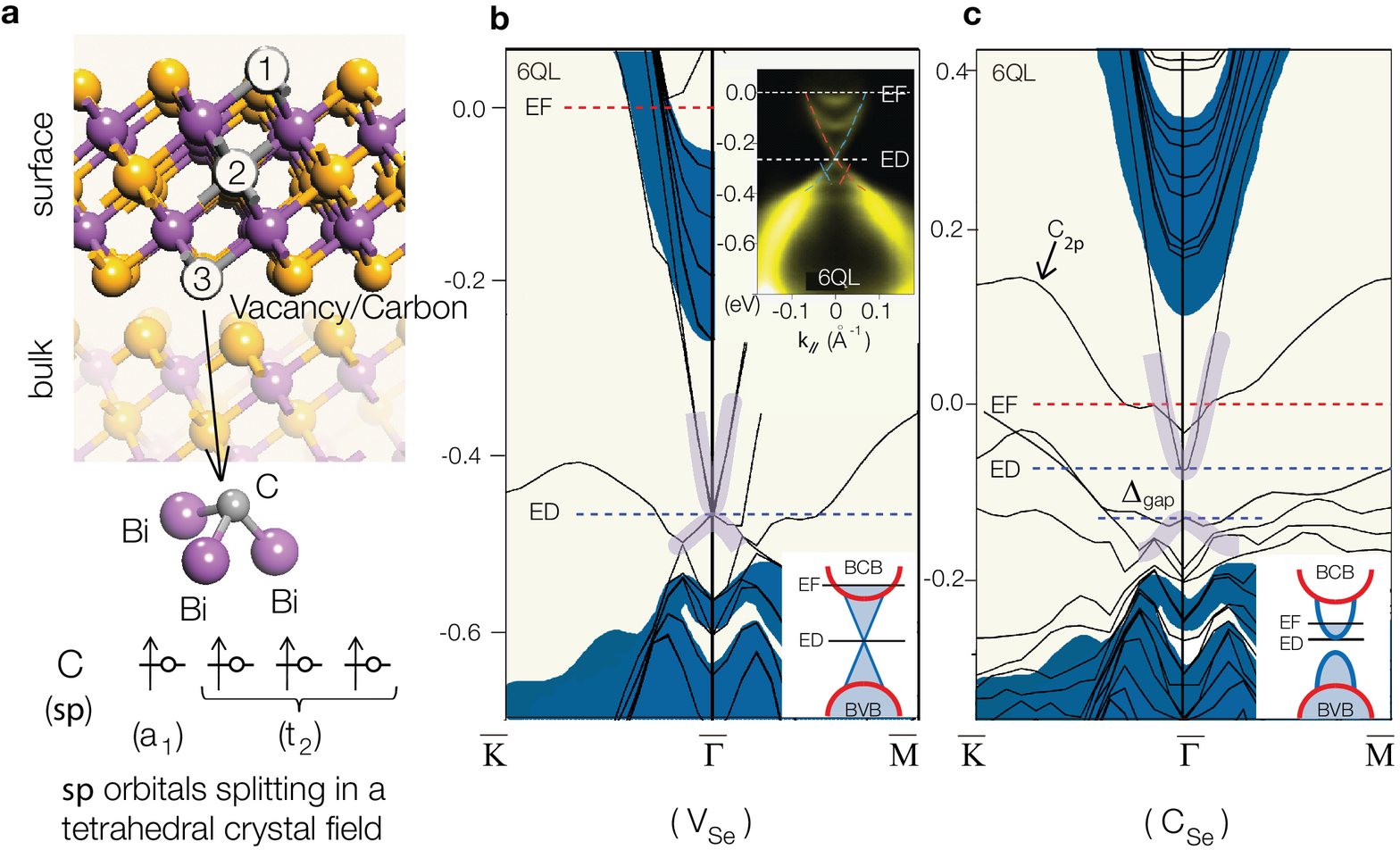}\\
\caption{(Color online) (a) Different anionic vacancy/doping sites on the surface quintuple layer of a Bi$_2$Se$_3$ slab, and schematic diagram of carbon substitutional doping at site 3 and splitting of its orbital under the tetrahedral crystal field. (b) Calculated band structure of 6QL Bi$_2$Se$_3$ thin films with a Se vacancy defect. The inset at the upper-right corner is the experimental band structure of 6QL Bi$_2$Se$_3$ thin films [ref.~39]. (c) Calculated band structure of 6QL Bi$_2$Se$_3$ thin films with one carbon substitutional doping. The projected pristine bulk band structure to the 2D Brillouin zone are shown in the background (blue). The surface topological states are highlighted by the bold purple lines. The insets (lower-right corners) are schematic illustrations of band structures of Bi$_2$Se$_3$ with Se vacancies (as-growth) and substitutional carbon doping at Se site (C$_{\rm Se}$), respectively. The Dirac energy level is labelled at the Dirac point without a gap and the bottom of the surface ``conduction" band with a gap, respectively.}
\end{figure}

\clearpage

\begin{figure}
\includegraphics[width=0.70\textwidth]{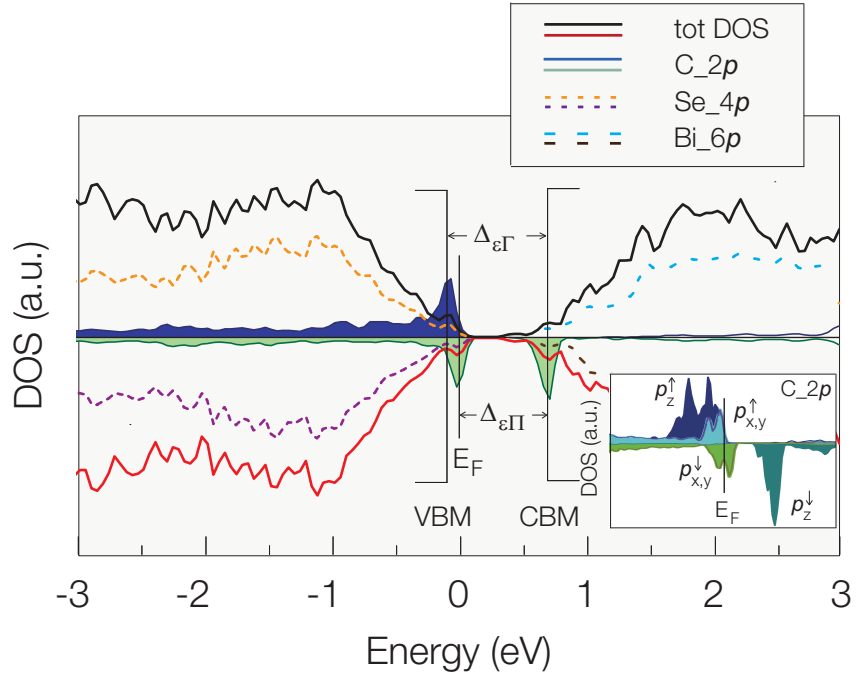}\\
\caption{(Color online) DOS (without SOI) of carbon doping in the Bi$_2$Se$_3$ thin film. The up panel is majority spin and the down panel is minority spin. The inset is the projected DOS on carbon 2$p$ orbitals.}
\end{figure}

\clearpage

\begin{figure}
\includegraphics[width=0.50\textwidth]{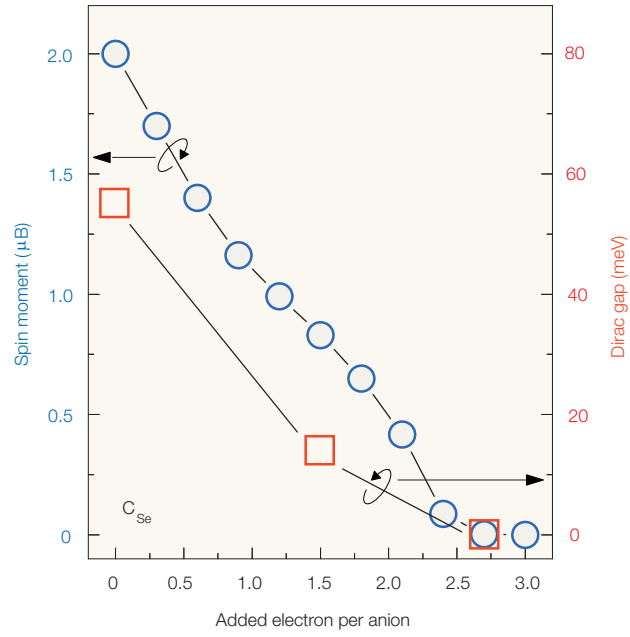}\\
\caption{(Color online) Dependency of spin moment and Dirac gap on the carrier (electron) concentration.}
\end{figure}

\clearpage

\begin{figure}
\includegraphics[width=0.9\textwidth]{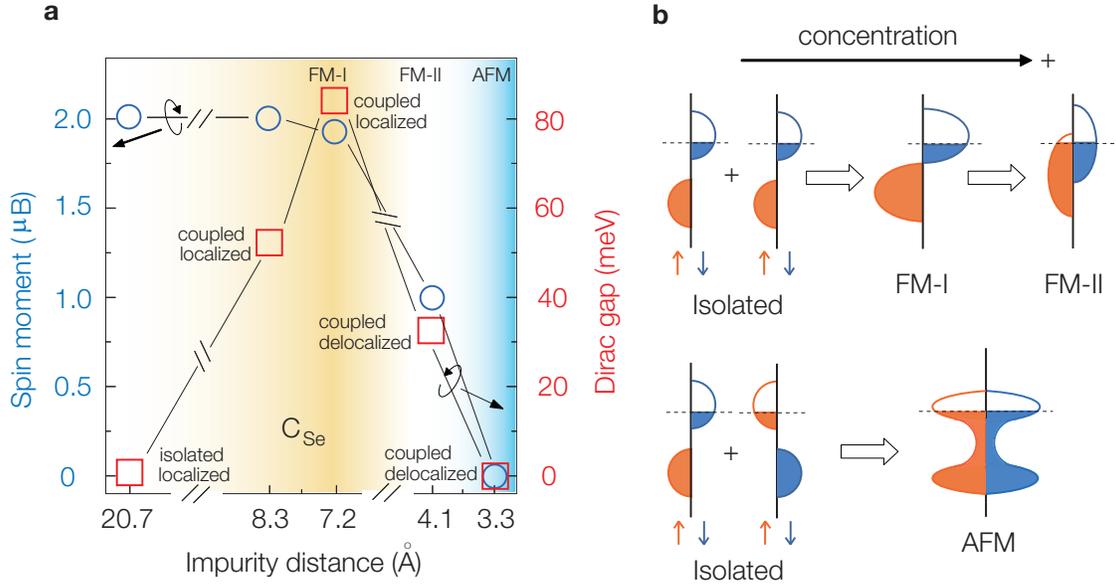}\\
\caption{(Color online).(a) Dependence of magnetic moment and Dirac gap on the inter-impurity distance, which is used to indicate impurity concentration, assuming a uniform impurity distribution. (b) Schematic illustration of the possible magnetic coupling in carbon doped TIs. The left figure shows the DOS for two isolated carbon dopants, and the right figure shows the coupling when the two carbon dopants are in close proximity at a higher doping concentration. The dashed lines denote the Fermi level.}
\end{figure}

\end{document}